\documentclass[letterpaper, 10 pt, conference]{ieeeconf}  

\usepackage{ragged2e}
\usepackage{subcaption}
\usepackage[T1]{fontenc}
\usepackage{graphicx}
\usepackage{booktabs}
\usepackage{algorithm}      
\usepackage{algorithmic}
\IEEEoverridecommandlockouts                              

\usepackage{graphics} 
\usepackage{amsmath} 
\usepackage{amssymb}  
\usepackage{mathtools}
\newcommand{\given}{\lvert}

\newcommand{\KL}{\text{KL}}
\newcommand{\E}{\mathrm{E}}
\DeclareMathOperator*{\argmax}{arg\,max}
\usepackage{array}
\newcolumntype{P}[1]{>{\centering\arraybackslash}p{#1}}
\title{\LARGE \bf
GOMA: Proactive Embodied Cooperative Communication via Goal-Oriented Mental Alignment
}

\author{Lance Ying$^{1,2}$, Kunal Jha$^{3}$, Shivam Aarya$^{4}$, Joshua B. Tenenbaum$^{2}$, Antonio Torralba$^{2}$, Tianmin Shu$^{4}$
\thanks{$^{1}$Harvard University, Cambridge, MA 02138, USA
        {\tt\small lanceying@seas.harvard.edu}}%
\thanks{$^{2}$Massachusetts Institute of Technology, Cambridge, MA 01239, USA
    {\tt\small jbt@mit.edu, torralba@csail.mit.edu}}%
\thanks{$^{3}$Dartmouth College, Hanover, NH 03755, USA
  {\tt\small kunal.a.jha.24@dartmouth.edu}}
\thanks{$^{4}$Johns Hopkins University, Baltimore, MD 21218, USA
    {\tt\small \{saarya1, tianmin.shu\}@mit.edu}}
}

\begin{document}

\maketitle
\thispagestyle{empty}
\pagestyle{empty}

\begin{abstract}

Verbal communication plays a crucial role in human cooperation, particularly when the partners only have incomplete information about the task, environment, and each other's mental state. In this paper, we propose a novel cooperative communication framework, Goal-Oriented Mental Alignment (GOMA). GOMA formulates verbal communication as a planning problem that minimizes the misalignment between the parts of agents' mental states that are relevant to the goals. This approach enables an embodied assistant to reason about when and how to proactively initialize communication with humans verbally using natural language to help achieve better cooperation. We evaluate our approach against strong baselines in two challenging environments, Overcooked (a multiplayer game) and VirtualHome (a household simulator). Our experimental results demonstrate that large language models struggle with generating meaningful communication that is grounded in the social and physical context. In contrast, our approach can successfully generate concise verbal communication for the embodied assistant to effectively boost the performance of the cooperation as well as human users' perception of the assistant.


\end{abstract}

\section{Introduction}

Rich verbal communication naturally emerges from human cooperation when people only have partial information about the environments and/or about each other's mental states \cite{tomasello2010origins}. It serves as a complementary source of information, in addition to the visual inputs, to help achieve better cooperation by aligning each other's mental states (including goals, beliefs, and eventually plans \cite{ying2023inferring, hadfield2016cooperative,gao2020joint}). Recent advances in large language models (LLM) and machine Theory of Mind (ToM) have sparked interest in building cooperative robots that can not only physically cooperate with humans but also verbally communicate with humans using natural language \cite{zhang2023building, hong2020multimodal}. However, it remains challenging to enable robots to actively \textit{initiate} verbal communication that is both concise (only communicate when necessary) and consistent with the physical environment and the social context (e.g., what humans want to do, believe, know, and need to know).

\begin{figure}[t!]
    \centering
    \includegraphics[width=0.47\textwidth]{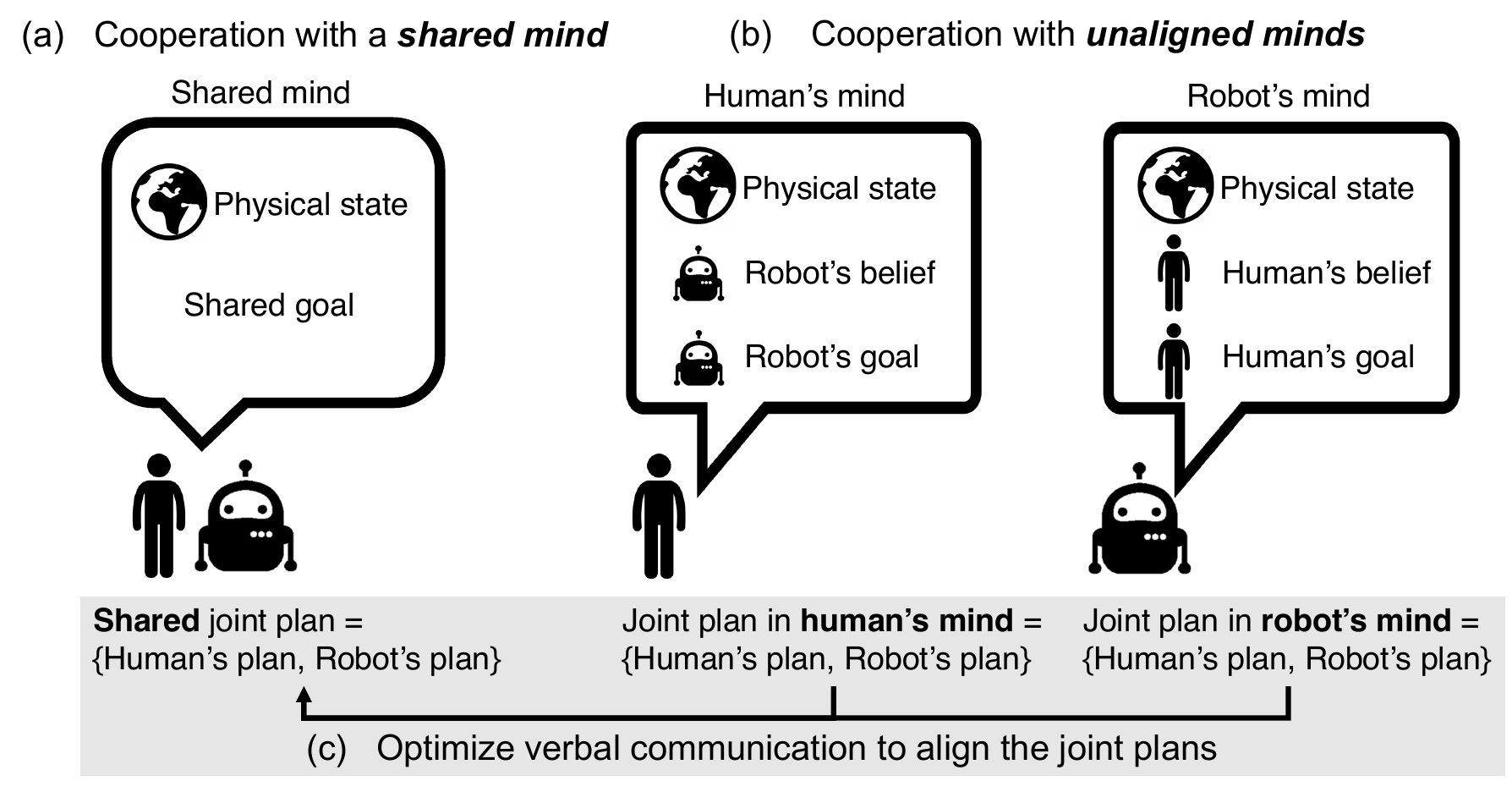}
    \caption{Illustration of cooperation with a shared mind or misaligned minds and communication optimized via goal-oriented mental alignment. (\textbf{a}) When human and robot minds are perfectly signed (i.e., a shared mind), they share the same belief of the physical state and the same goal, which leads to the same joint plan shared by both agents, an ideal condition for human-robot cooperation. (\textbf{b} However, in real-world tasks, human and robot minds are typically unaligned, leading to two different (often conflicting) joint plans in their minds. (\textbf{c}) To achieve a shared joint plan that optimizes cooperation, we optimize the robot's verbal communication to actively align the joint plans in both agents' minds.}
    \label{fig:intro}
\end{figure}



Studies in psychology has shown that proactive verbal communication serves to align the mental states of agents \cite{clark1996using}. Imagine you are going to get some groceries for your mom. As you put on your shoes and walk towards the door, your mom gets out of the kitchen and says ``It's going to rain, get your umbrella, and don't forget about the avocados.'' In this scenario, your mom decides to communicate with you because she is uncertain whether you have the same beliefs regarding the weather forecast and the required grocery items as you walk out the door. 


When cooperating with one another, each agent not only needs to plan for itself but also has to imagine the plans of its partners. Such planning process is termed as \textit{joint planning} \cite{kleiman2016coordinate, wu2021too}. To achieve joint planning, prior works typically assumed that both agents have full observability and complete knowledge about the task. In other words, they have a shared mind, based on which they can derive the same joint plan (Fig.~\ref{fig:intro}(\textbf{a})). However, in real-world embodied cooperation, robot assistants only have partial observations and often do not know the true human goals (Fig.~\ref{fig:intro}(\textbf{b})).

The goal of cooperative communication is then to reach a shared mind (two agents' are perfectly aligned) so that the resulting joint plans in both agents' minds are the same (Fig.~\ref{fig:intro}(\textbf{c})). Once we reach such mental alignment condition, both agents know exactly what each other plans to do, and therefore achieve optimal cooperation. However, an agent belief can be about any part of a state. If the state is high dimensional, it is extremely difficult to make sure two beliefs are the same. Our key insight is that we \texttt{only} need to align the part of the belief that is relevant to reaching the goal.

Following this insight, we propose a novel cooperative communication framework, Goal-Oriented Mental Alignment (GOMA). In this framework, we aim to generate optimal communication in the belief space. That is, verbal communication, by exchanging information, can help reshape agents' beliefs. In particular, GOMA first seeks to detect misalignment in agents' goal-relevant beliefs via divergence between the joint plans based on an agent's own belief and a simulated hypothetical mind after acquiring additional knowledge from another agent via communication. We then optimize the communication using a proxy reward derived from the divergence between the plans. The resulting communication can then minimize the difference between the joint plan in each agent's mind and the true joint plan, which improves agents' coordination.

We evaluate GOMA in two popular human-AI cooperation domains, Overcooked and VirtualHome. Our experimental results with a simulated human agent and real human participants show that our GOMA outperforms strong baselines (including a recent LLM-based baseline). The GOMA-enabled assistant also receives higher subjective ratings from human participants.

In sum, our contributions include (1) a novel embodied cooperative communication framework -- GOMA, (2) extensive evaluation of strong baselines and GOMA in two challenging domains, and (3) a human user study that evaluates the task performance of AI assistants and humans' perception of them.

\section{Related Work}
\subsection{Communication in Collaboration}
Human communication is grounded in cooperative intentions. \cite{clark1996using} argues that language communication is a joint activity that attempts to achieve mutual understanding. \cite{tomasello2010origins} proposes three communicative motives: requesting help or information, informing the other agents, and sharing feelings or attitudes. These communicative motives help to align the mental states of the agents. Through verbal communication, agents can assess others' goals, knowledge, emotions, and beliefs, which they can then use to plan for the next actions.

However, verbal communication can also be costly, as it demands cognitive resources and distracts agents when performing actions \cite{macmillan2004communication}. Prior works on multi-agent teaming have formalized communication costs in collaborative settings \cite{macmillan2004communication, horvitz2003learning, unhelkar2020decision}, showing that excessive communication can degrade the performance of the team. Therefore, when designing communication policies, the AI assistant needs to communicate useful, concise, and relevant information yet not too frequently.

\subsection{Collaborative and Communicative AI Agent}

Communication between humans and robots has also been extensively studied. Most existing literature has focused on one-directional communication where the human instructs the robot \cite{williams2018learning, zhi2024pragmatic}. Some recent studies have proposed bi-directional communication. For example, Yuan et al. \cite{yuan2022situ} proposes a bi-directional human-robot collaborative communication framework that allows the robot to communicate decisions with explanations from human feedback. Unhelkar et al. \cite{unhelkar2020decision} introduced CommPlann, a bi-directional communication framework that allows the robot to ask for human's intent, share the robot's intent, and give commands to humans. There have been recent works that use LLMs as a communication module in bi-directional human-robot communication, (e.g., \cite{zhang2023building,zhang2023large, ahn2022can, mandi2023roco}). While these recent LLM-based agents can achieve certain success, the communication generated by LLMs is often redundant and/or not grounded in agents' mental states, actions, and plans.

In addition, most human-robot communication frameworks, such as \cite{unhelkar2020decision,devin2016implemented, schaefer2017communicating}, assume full agent observability. The resulting communication is thus only restricted to informing and inquiring about goals and plans. Our work extends to scenarios where both agents have partial observability of the environment and allows the robot to communicate to resolve partial knowledge and false beliefs about the environment state. This requires agents to model and reason about each other's mental states recursively (e.g., the robot thinks the human thinks the glass is in the fridge, but it knows that it is in the cabinet), which remains a challenge for LLMs today \cite{ullman2023large}. As a result, such cooperative communication capacity remains an open research question in embodied cooperation.

\subsection{Theory of Mind for Cooperative Robot Planning}

There have been many studies on inferring an agent's goals and beliefs (e.g., \cite{baker2017rational, zhi2020online, shu2021agent, jin2024mmtom, ying2023neuro, ying2023inferring, ying2024grounding}), commonly referred to as Theory of Mind reasoning, to better coordinate with humans in collaborative tasks. Previous studies have leveraged explicit mental reasoning to improve cooperative robot planning. This includes generating more expressive or explainable plans to improve humans' understanding of robots' plans \cite{dragan2013generating, stulp2015facilitating,kwon2018expressing, zhang2017plan,gao2020joint,gao2022show} or better understanding of humans' cooperative actions \cite{hadfield2016cooperative}, all via reasoning about humans' mental models of the robot. There have also been works on developing a shared joint planner in two agents' minds to reach optimal coordination by reasoning about both agents' plans jointly. However, existing works do not allow verbal communication between humans and robots in addition to action planning. Our work aims to fill this gap by jointly planning for actions that change the physical state and verbal communication that changes the mental states of humans and robots.

\section{Problem Formulation}\label{sec:formulation}

In this work, we consider two agents, a human user and a robot assistant. To successfully communicate and cooperate, the two agents must infer each other's mind. We adopt the Interactive Partially Observable Markov Decision Process (I-POMDP) \cite{gmytrasiewicz2005framework, doshi2009monte} to formulate the mental reasoning between the human and the robot.

\subsection{Background: I-POMDP}

I-POMDP is a framework that enables an agent to recursively model other agents, which captures complex social interactions between agents. Here, we consider the interactions between two agents, $i$ and $j$, in which agent $i$ infers agent $j$'s mental state recursively. In an I-POMDP, there are states $s^t$; agents' observations, $o_i^t$ and $o_j^t$, sampled from their conditional observation probabilities, $O_i(o^t_i | s^t)$ and $O_j(o^t_j | s^t)$; and agents' actions $a_i^t$ and  $a_j^t$. Agents have their beliefs, $b_i^t$ and $b_j^t$, and goals, $g_i$ and $g_j$. To model the recursive mental reasoning, we define interactive states for the agents, i.e., $is_{i, \ell}$ and $is_{j, \ell}$,  at level-$\ell$. From agent $i$'s perspective, we define its interactive state at each level as
\begin{itemize}
    \item Level $0$: $is_{i, 0} = s$
    \item Level $1$: $is_{i, 1} = (s, b_{j, 0}, g_j)$ 
    \item $\cdots$
    \item Level $\ell$: $is_{i, \ell} = (s, b_{j, \ell - 1}, g_j)$ 
\end{itemize}

The level-$\ell$ inference for agent $i$ is to infer the belief $b_{i, \ell}^t = p(is_{i, \ell}^t \given o_i^{1:t}, a_i^{1:t - 1})$. Since the level-$\ell$ agent $i$'s interactive state, $is_{i, \ell}^t = (s^t, b_{j, \ell - 1}^t, g_j)$, includes $j$'s belief at level $\ell - 1$ ($b_{j, \ell - 1}^t$), the inference at level $\ell$ depends on inference at level $\ell - 1$ which depends on inference at level $\ell - 2$, and so on. This recursive inference terminates at level $0$. That is, the belief at level-0 is only about the physical state, $b_{i, 0} = p(s^t \given o_i^{1:t}, a_i^{1:t - 1})$. This becomes a standard POMDP \cite{kaelbling1998planning} which does not model other agents.

\subsection{Two-level Reasoning for Embodied Cooperative Communication}\label{sec:formulation_embodied}

Theoretically, the level of agents' reasoning about other agents' minds can go to infinity (e.g. robot thinks human thinks robot thinks...) yet we cap the depth at two in our model, which is in line with most empirical evidence suggesting that humans rarely engage in greater than 2 levels of recursive Theory of Mind reasoning \cite{bosch2002one}. Therefore, we adopt a two-level I-POMDP for modeling the mental reasoning between a human user and a robot assistant in embodied cooperation. In particular, we define the mind of each agent as the belief of the level-1 interactive state of the agent. For the human user's mind, we have $m_H = (b(is_{H,1}), g_H) = ((b_{H,0}, b(b_{R,0}), b(g_R)), g_H)$, where $b_{R,0}$ is the robot's interactive state at level 0, i.e., its belief about the physical state; $g_R$ is the robot's goal; and $g_H$ is the human's goal. Similarly, for the robot assistant, we define its mind as $m_R = (b(is_{R,1}), g_R) = ((b_{R,0}, b(b_{H,0}), b(g_H)), g_R)$, where $b_{H,0}$ is the human's belief about the physical state. Intuitively, each mind models the agent's belief about (1) the physical state, (2) another agent's belief about the physical state, and (3) the goal of another agent. Due to the cooperative nature of our problem setting, we further constrain the goal inference to be either one of the following two conditions:
\begin{itemize}
    \item Condition 1: Both agents share a known common goal;
    \item Condition 2: The robot's goal is the human goal inferred by the robot, and the human user knows that the robot is trying to help with the inferred human goal.
\end{itemize}



Condition 1 models human-robot teaming, in which the human and robot agents are teammates who work on the same task assigned to them a priori. Condition 2 models robot assistance, in which the human's true goal is unknown to the robot a priori, thus the robot must infer the human's goal and provide assistance. In both cases, agents only have partial observability of the physical state, and thus they have to infer both the physical state and each other's belief about the physical state. It is worth noting that our formulation departs from most previous assistance-game setups, which either assume that the agents have full observability or that they share a known goal. As in collaborative tasks, agents often do not have perfect knowledge of the environment and thus need to represent other agents' beliefs differently from theirs and communicate and coordinate their actions, our formulation is more aligned with real-world embodied cooperation.


\section{Goal-Oriented Mental Alignment}



As Fig.~\ref{fig:intro} illustrates, when there is a shared mind, two agents will share the same joint plan. In our Goal-oriented Mental Alignment (GOMA) framework, we formulate communication optimization as the convergence of the current joint plan and the joint plan given a shared mind achieved by exchanging information through verbal communication. In particular, we consider two types of communication -- sharing information and requesting information. These are two dominant types of verbal communication in human cooperation \cite{tomasello2010origins}. We hypothesize that these are also two types of communication that a robot assistant can \textit{proactively} initiate to achieve joint plan alignment. To reason whether to communicate and what to communicate, we define a proxy reward for minimizing the divergence between plans before and after one type of communication. We summarize GOMA in Algorithm~\ref{alg:GOMA}, which works with any off-the-shelf action planner. We introduce key components of the algorithm in the rest of the section.

\begin{algorithm}[t!]
\footnotesize
   \caption{GOMA}
   \label{alg:GOMA}
\begin{algorithmic}[1]
   \STATE {\bfseries Input:} $\textbf{Planner()}$, $T_\text{max}$
   \STATE {\bfseries Initialization:} $b(g_H)$, $b_{R, 0}(s^0)$, particles of sampled human beliefs: $\{b_{H,0}^{(l)}(s^0)\}_{l=1}^L$
   \STATE $t \leftarrow 1$, $u_R^0 = \text{None}$
   \REPEAT 
   \STATE Observe $o_R^t$ and receive human message $u_H^{t-1}$
   \STATE Update level-0 belief: $b_{R, 0}(s^t)$ based on both $o_R^t$ and $u_H^{t-1}$
   \STATE Robot knowledge: $K_R^t = K_R(b_{R,0}(s^t))$
   \STATE Update human goal inference: 
   \STATE $b(g_H) \propto P(a_H^{t-1} | g_H) P(u_H^{t-1} | g_H) b(g_H), \forall g_H \in \mathcal{G}$
   \FORALL{$l = 1,\cdots,L$} 
     \STATE Sample a human goal based on the goal inference: $\hat{g}^{(l)}_H \sim b(g_H)$
     \STATE Set the robot goal as the inferred human goal: $g_R^{(l)}\leftarrow \hat{g}^{(l)}_H$
     \STATE Sample an environment state $s^t \sim b_{R, 0}(s^t)$
     \STATE Sample inferred human observations $\hat{o}_H^t \sim O_H(\hat{o}_H^t| s^t)$
     \STATE Update $b_{H, 0}^{(l)}(s^t)$ based on both $\hat{o}_H^t$ and $u_R^{t-1}$
     \STATE Human plan given the inferred human belief:
     \STATE $\pi_H(a_H^t | b_{H,0}, \hat{g}^{(l)}_H) \leftarrow \textbf{Planner}(b_{H,0}, \hat{g}^{(l)}_H)$
     \STATE Human plans given the shared minds augmented by different sub-states in robot knowledge:
     \STATE $\{\pi_H(a_H^t | b^{+s_n}_{H,0}, \hat{g}^{(l)}_H) \leftarrow \textbf{Planner}(b^{+s_n}_{H,0}, \hat{g}^{(l)}_H)$; $\forall s_n^t \in K_R^t\}$
     \STATE Robot plan given the robot belief:
     \STATE $\pi_R(a_R^t | b_{R,0}, g_R^{(l)}) \leftarrow \textbf{Planner}(b_{R,0}, g_R^{(l)})$
     \STATE Robot plans given the shared minds augmented by different sub-states in human knowledge:
     \STATE $\{\pi_R(a_R^t | b^{+s_n}_{R,0}, g_R^{(l)}) \leftarrow \textbf{Planner}(b^{+s_n}_{R,0}, g_R^{(l)})$; $\forall s_n^t \in K(b_{H,0}^{(l)}(s^t))\}$
    \ENDFOR
    \STATE $m_R \leftarrow (b_{R,0}(s^t), \{b_{H,0}^{(l)}(s^t)\}_{l=1}^L, b(g_H))$
    \STATE All possible human knowledge: $\hat{K}_H^t = \cup_{l=1}^L K(b_{H,0}^{(l)}(s^t))$
    \STATE Construct the utterance space $U$ based on the robot knowledge $K_R^t$ and all possible kuman knowledge $\hat{K}_H^t$
    \STATE Compute $R(u, M_R)$, $\forall u \in U$ using the plans generated above based on Eq.~(\ref{eq:reward_1}-\ref{eq:reward_3})
    \STATE Select robot utterance based on the proxy reward:
    \STATE $u_R^t = \argmax_{u\in U} R(u, M_R)$
    \STATE Select robot action based on the average plan:
    \STATE $a_R^t =\argmax_{a_R \in \mathcal{A_R}} \sum_{l=1}^L \pi_R(a_R | b_{R,0}, g^{(l)}_R) / L$
    \STATE Execute the robot action $a_R^t$ and send the robot utterance $u_R^t$
    \STATE $t \leftarrow t + 1$
   \UNTIL $t = T_\text{max}$ or the true goal has not been reached
\end{algorithmic}
\end{algorithm}

\subsection{Goal Inference and Joint Planning for the Robot}

Unless the human goal is given to the robot a priori (i.e., condition 1 defined in Section~\ref{sec:formulation_embodied}), the robot must infer the human goal. We adopt the approach introduced by \cite{ying2023inferring}, which leverages an LLM to conduct goal inference based on the observed human actions and messages (Line 8-9 in Algorithm~\ref{alg:GOMA}). We then sample the possible goals of humans 

The joint plan for the robot includes two components. First, the robot's policy given its goal and its belief, i.e., $\pi_R(a_R | b_{R,0}, g_R)$. Second, the expected human's policy inferred by the robot, i.e., $\E_{b(b_{H,0}),b(g_H)}[\pi_H(a_H | b_{H,0}, g_H)]$. In practice, we can estimate this expectation via sampling particles of possible human beliefs (i.e., $\{b_{H,0}^{(l)}(s^t)\}_{l=1}^L$ in Algorithm~\ref{alg:GOMA}) and possible goals (Line 11 in Algorithm~\ref{alg:GOMA}). 

\subsection{Agent Knowledge From Level-0 Belief}

Recall that the level-0 belief of an agent $b_{i,0}$ represents the agent's belief of the physical state $s$. If we partition the state $s$ into multiple sub-states such as states of all objects in the environment, then we can evaluate the uncertainty in the belief of each sub-states. We define the sub-states that have certain belief distributions as knowledge of an agent. Formally, let us denote a state partition as $s = \{s_n\}_{n=1}^N$ with N sub-states and $b_{i,0}(s_n)$ as the level-0 belief of the sub-state $s_in$. For instance, if $s_n$ is object $n$'s state, then $b_{i,0}(s_n)$ is the belief of the object $i$' state. Consequently, we define the knowledge of agent $i$ as 
\begin{align}
K_i &= K(b_{i,0}) \nonumber\\
& = \{b_{i,0}(s_n) : \mathcal{H}(b_{i,0}(s_n)) < \mathcal{H}_\text{max}, n=1,\cdots,N\},    
\end{align}
where $\mathcal{H}$ is the entropy of a belief distribution and $\mathcal{H}_\text{max}$ is maximum entropy that is considered to be certain. In the example of object states as sub-states, knowledge consists of objects over which the agent has beliefs with high certainty. 

\subsection{Shared Mind Augmented by An Agent's Knowledge}

An agent $i$ can imagine a shared mind after acquiring knowledge about a sub-state, $b_{j,0}(s_n) \in K_i$, from another agent $j$ via verbal communication, as both agents would share this knowledge after the communication. We define this as the belief merge operation $b^{+s_n}_{i,0} = \text{Merge}(b_{i,0}, b_{j,0}(s_n))$. Specifically, this merge operation will set the belief of sub-state $n$ of agent $i$ to that of agent $j$, i.e., $b^{+s_n}_{i,0}(s_n) = b_{j,0}(s_n)$.

\subsection{Divergence Between Plans as Proxy Reward}

It is hard to directly estimate the effect of an utterance on the overall task performance. To directly reason what knowledge is critical for aligning the joint plans between agents, we define a proxy reward for communicating about the knowledge of an agent's knowledge. Since the goal of this work is to generate proactive communication initiated by the robot, we model the proxy reward from the perspective of the robot.

We first define the reward of sharing the robot's knowledge of sub-state $s_n$ with the human user as the Kullback–Leibler (KL) divergence between the human agent's plan after communication and before communication.
\begin{align}\label{eq:reward_1}
    &R(\text{share } s_n, M_R) =\nonumber\\
    & \KL\left(\E[\pi_H(a_H | b^{+s_n}_{H,0}, g_H)] || \E[\pi_H(a_H | b_{H,0}, g_H)]\right) - C,
\end{align}
where $b^{+s_n}_{H,0} = \text{Merge}(b_{H,0}, b_{R,0}(s_n))$ and $C$ is the cost for communication at a time step.

We then define the reward of requesting possible human knowledge of sub-state $s_n$ to inform the robot's plan:
\begin{align}\label{eq:reward_2}
    &R(\text{request } s_n, M_R) =\nonumber\\
    & \KL\left(\pi_R(a_R | b^{+s_n}_{R,0}, g_R) || \pi_R(a_R | b_{R,0}, g_R)\right) - C,
\end{align}
where $b^{+s_n}_{R,0} = \text{Merge}(b_{R,0}, b_{H,0}(s_n))$.

The plans used to compute the KL-divergence for the proxy rewards can be generated by running an off-the-shelf planner given the corresponding beliefs and goals (Line 16-23 in Algorithm~\ref{alg:GOMA}).

We also define the reward for not communicating at a step as follows:
\begin{equation}\label{eq:reward_3}
    R(\text{None}, M_R) = 0.
\end{equation}

\subsection{Communication Optimization}

Given the proxy rewards defined above, we can then choose whether and what to communicate based on the robot's mind at each step (Line 27-30 in Algorithm~\ref{alg:GOMA}). In particular, the utterance space is $U = \{\text{None}\} \cup \{\text{share } s_n; s_n \in K_R\} \cup \{\text{request } s_n; s \in \hat{K}_H\}$, where $\hat{K}_H$ is the inferred human knowledge estimated from the human belief particles: $\hat{K}_H = \cup_{l=1}^L K(b_{H,0}^{(l)})$. We select the best robot utterance at step $t$ as follows:
\begin{equation}
    u_R^t = \argmax_{u\in U} R(u, M_R).
\end{equation}

We can further generate a natural language message based on the utterance $u_R^t$ to enable communication with real humans. This can be achieved by using GPT-4 \cite{openai2023gpt4} to translate $u_R^t$ to natural language through few-show prompting.

\subsection{Multimodal Mental Update}
At each step, the robot will update its mind based on both its observation $o^t_R$ and the messages it sends and receives. In particular, we extract human knowledge $b_{H,0}(s_n)$ from the human message $u_h^t$ via GPT-4 and use it to update the robot's level-0 belief $b_{R,0}$ jointly with $o^t_R$ (Line 6 in Algorithm~\ref{alg:GOMA}) operation. For instance, if the human informs the robot of the location of an object, we can update the robot's level-0 belief with the knowledge of the object's location. Additionally, if the robot shares knowledge $b_{R,0}(s_n)$ in its utterance, then the robot can assume that the human's level-0 belief will also be updated accordingly. Thus, in robot mind $M_R$, we can update $b(b_{H,0})$ using both the shared robot knowledge and the human observation (Line 15 in Algorithm~\ref{alg:GOMA}). Note that we can sample possible human observations based on the state inferred by the robot's level-0 belief (Line 13-14 in Algorithm~\ref{alg:GOMA}). All beliefs are initialized with a uniform distribution (Line 2 in Algorithm~\ref{alg:GOMA}).

\section{Experiments}

We evaluate our model in two human-AI domains Overcooked and VirtualHome. These two domains cover two distinct alignment objectives. In both domains, there are two agents -- a human user and an embodied AI assistant. In Overcooked, the agent's goal is to align their plans temporally so that certain joint actions can be performed at similar time steps, whereas in VirtualHome, the agents align their beliefs about the location of the objects they try to collect.  We describe each in detail below.

\begin{table}[t]
    \centering
\begin{tabular}{cP{6cm}}
\toprule
\textbf{Recipe Name} & \textbf{Ingredient List} \\
\midrule
Burger & Cooked(Patty), Cooked(Potato), Chopped(Lettuce), Chopped(Tomato)\\
\hline
Pasta & Cooked(Spaghetti), Cooked(Mushroom), Cooked(Cream), Chopped(Basil) \\
\hline
Ramen & Cooked(Noodle), Cooked(Mushroom), Cooked(Egg), Chopped(Scallion) \\
\hline
Steak \& Fries & Cooked(Beef), Cooked(Potato), Chopped(Parsley) \\
\bottomrule
\end{tabular}
    \caption{Overcooked recipe specifications.}
    \label{tab:oc-goal}
\end{table}


\begin{figure}[t!]
    \centering
    \includegraphics[width = 0.25\textwidth]{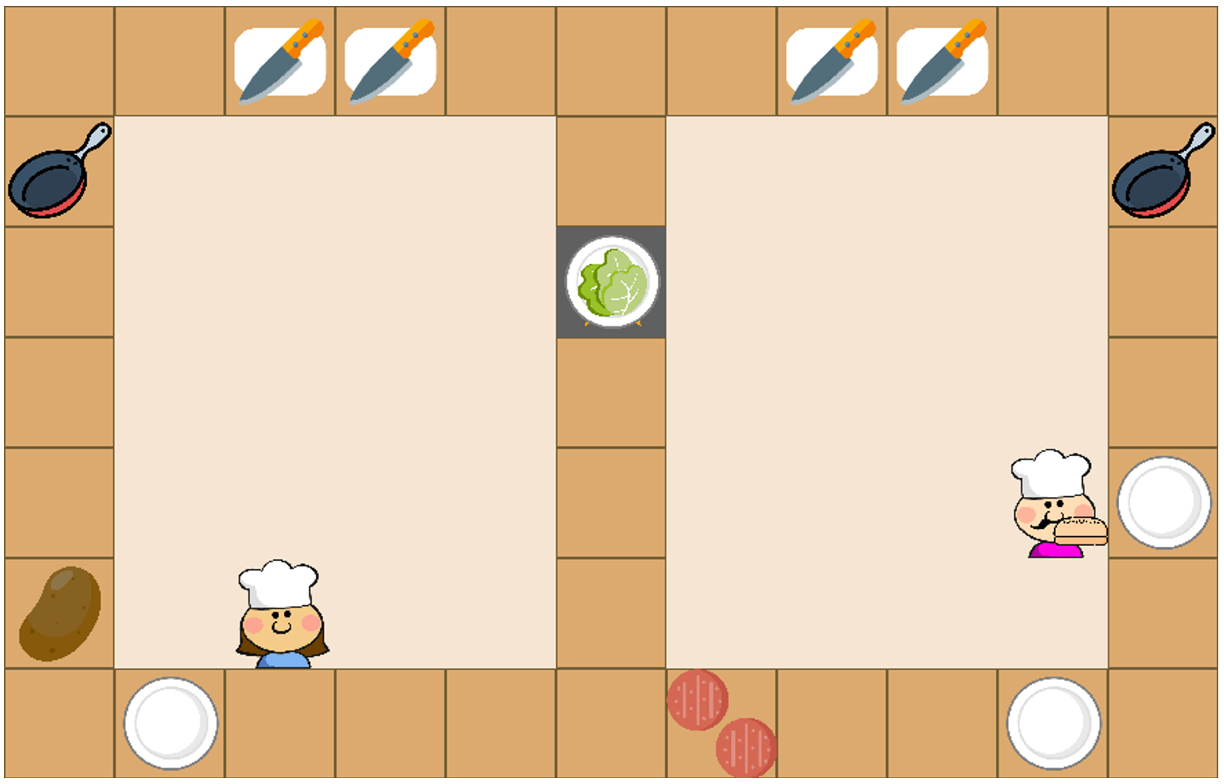}
    \caption{Example Overcooked environment. In each environment, there are two rooms. The two agents are always in different rooms. An agent cannot observe the other room and has to rely on verbal communication to infer the states of the objects in the other room.}
    \label{fig:overcooked_env}
    \vspace{-5pt}
\end{figure}

\subsection{Overcooked}
Overcooked is a popular multiagent game where agents need to collaborate to prepare and cook ingredients, which is also widely used for evaluating human-AI cooperation (e.g., \cite{carroll2019utility, wu2021too}). In the original game, agents have full observability. In this study, we extended the Overcooked simulator from \cite{wu2021too} by assuming partial observability where each agent cannot observe the other room as shown in Fig.~\ref{fig:overcooked_env}. At each step, the AI assistant may share its progress on the task or ask about the human's progress.

The goal of the collaborating agents is to complete the dishes in the shortest amount of time. To simulate more realistic cooking scenarios, we augment the existing simulator with dynamics that cooked ingredients will gradually cool down. If cooked ingredients are not at the ideal temperature when the dish is served, the team will receive a penalty. This requires both agents to coordinate to avoid temporal misalignment in their plans. For instance, one agent should not finish making the burger too early if the other agent has not started cooking the French fries. The agents can align their plan by choosing to wait for the other agents (e.g. I will start cooking A as soon as the other agent finishes B). There are four recipes in our experiment (Table~\ref{tab:oc-goal}): Burger, Spaghetti, Ramen, and Steak, each in a unique room layout. We simulate a human agent using the planner in \cite{wu2021too}. Each recipe is run 10 times with different seeds and we report the aggregate results.

\textbf{Baselines.} We evaluate three baselines: Single-agent, No-Communication (No-Comm), and Heuristic-based Communication (Heur-Comm). In Single-Agent, the human completes all the tasks alone. In the No-Comm baseline, no messages are exchanged. In the Heur-Comm baseline, the AI Assistant follows a simple heuristic that shares updates every time a sub-goal has been completed and periodically asks for the human's progress. The action planner in all methods including GOMA is the same as the planner in \cite{wu2021too}.

\textbf{Metrics.}
We use two performance metrics: speedup and total plan costs. Speedup is calculated by comparing the plan length in each team condition, where the human is working with one of the four collaborative AI models, to the single agent baseline, i.e. $ Speedup = L_{single}/L_{team} - 1$.

Total plan cost is the sum of all action and communication costs with penalties applied for sub-optimal dish states due to time lapse between the completion of a hot sub-task (e.g. cooked noodle) and the end of the trial, i.e. $Total Cost = L + U + \sum_{i \in hot\_items}{\Delta(L_i, L)}$ where $U$ is the total number of utterances in a trial, $L$ is the plan length and $L_i$ refers to the time step where item i is completed.

\begin{table}[t]
    \centering
\begin{tabular}{P{2cm}P{6cm}}
\toprule
\textbf{Goals} & \textbf{Goal Specification} \\
\midrule
Set up table & Put [N forks, N plates, N waterglasses or wineglasses] on [kitchentable, coffeetable]\\
\hline
Put groceries & Put [N apple, N salmon, N pudding, N cupcakes] inside [cabinet, fridge]\\
\hline
Prepare food & Put [N apple, N salmon, N pudding, N cupcakes] on [kitchentable, coffeetable] \\
\hline
Load dishwasher & Put [N forks, N plates, N waterglasses or wineglasses] inside [dishwasher] \\
\bottomrule
\end{tabular}
    \caption{VirtualHome goal specifications.}
    \label{tab:vh-goal}
    \vspace{-10pt}
\end{table}

\begin{figure*}[htb!]
    \begin{subfigure}[t]{0.48\textwidth}
    
        \centering
                \caption{Overcooked Simulation}
        \includegraphics[width = \textwidth]{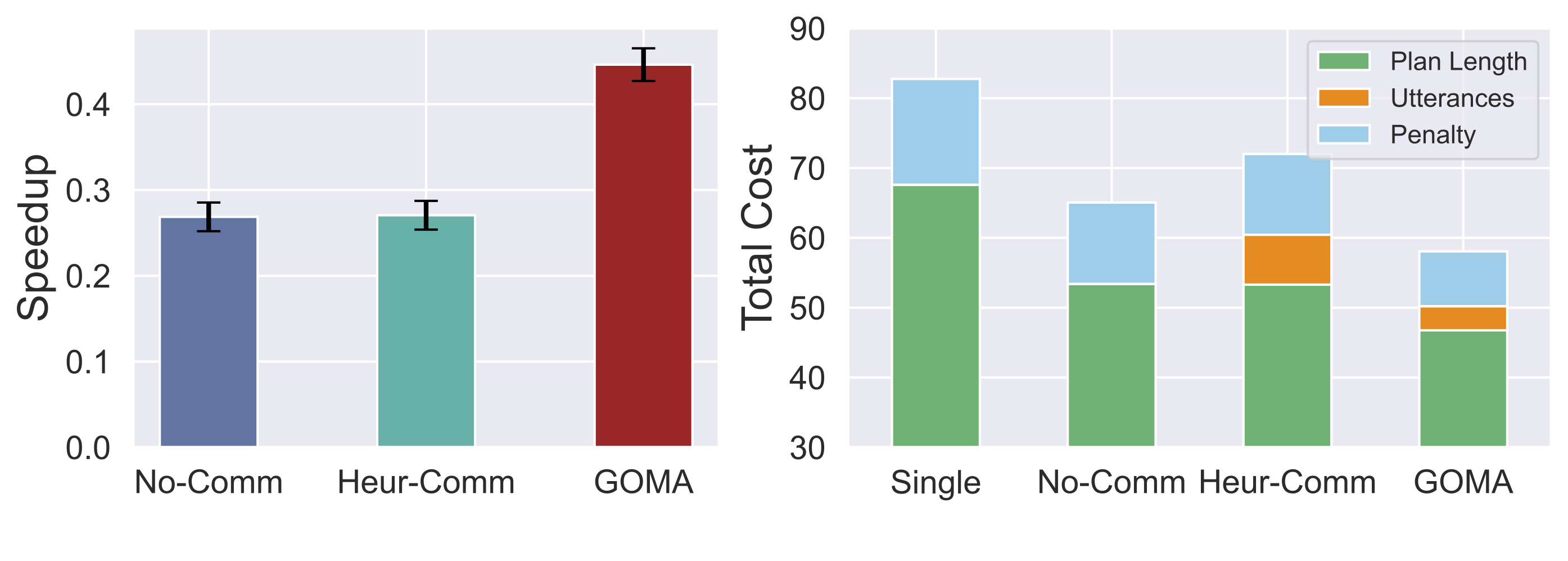}

    \end{subfigure}%
    ~ 
    \begin{subfigure}[t]{0.48\textwidth}
        \centering
            \caption{Virtual-Home Simulation}
        \includegraphics[width = \textwidth]{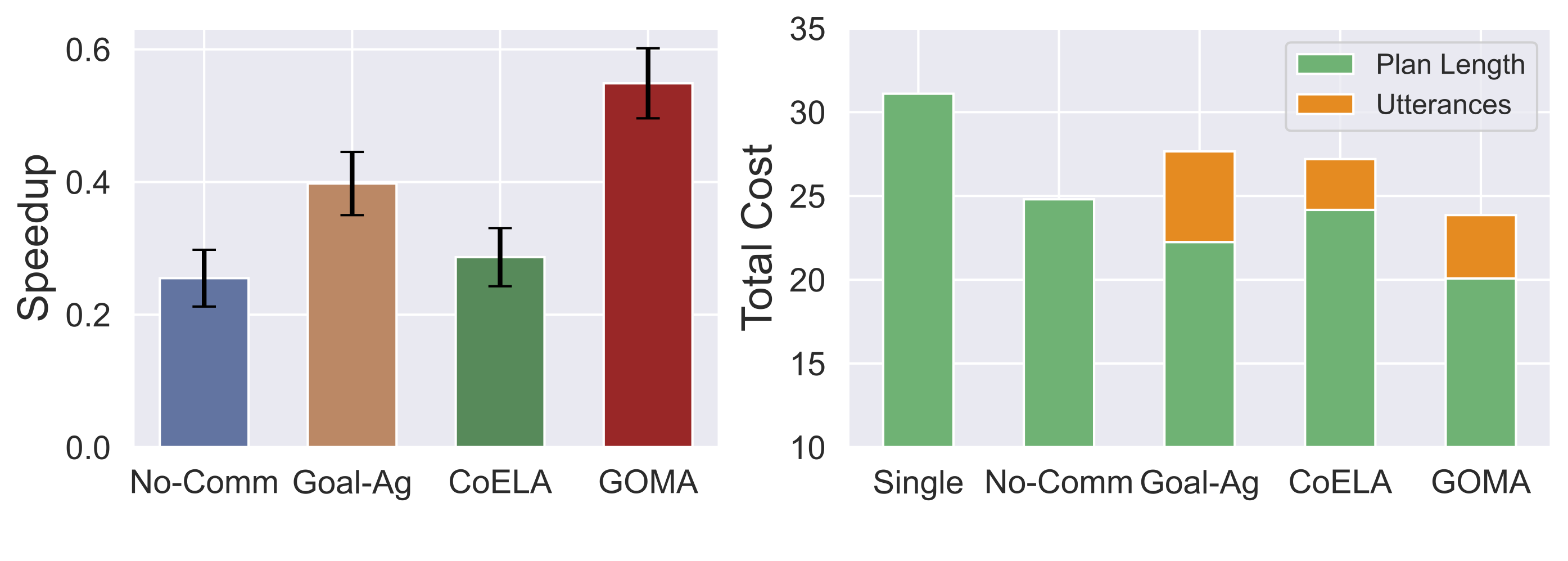}

    \end{subfigure}
    
    \begin{subfigure}[t]{0.48\textwidth}
        \caption{Virtual-Home Human Experiment}
        \centering
        \includegraphics[width = \textwidth]{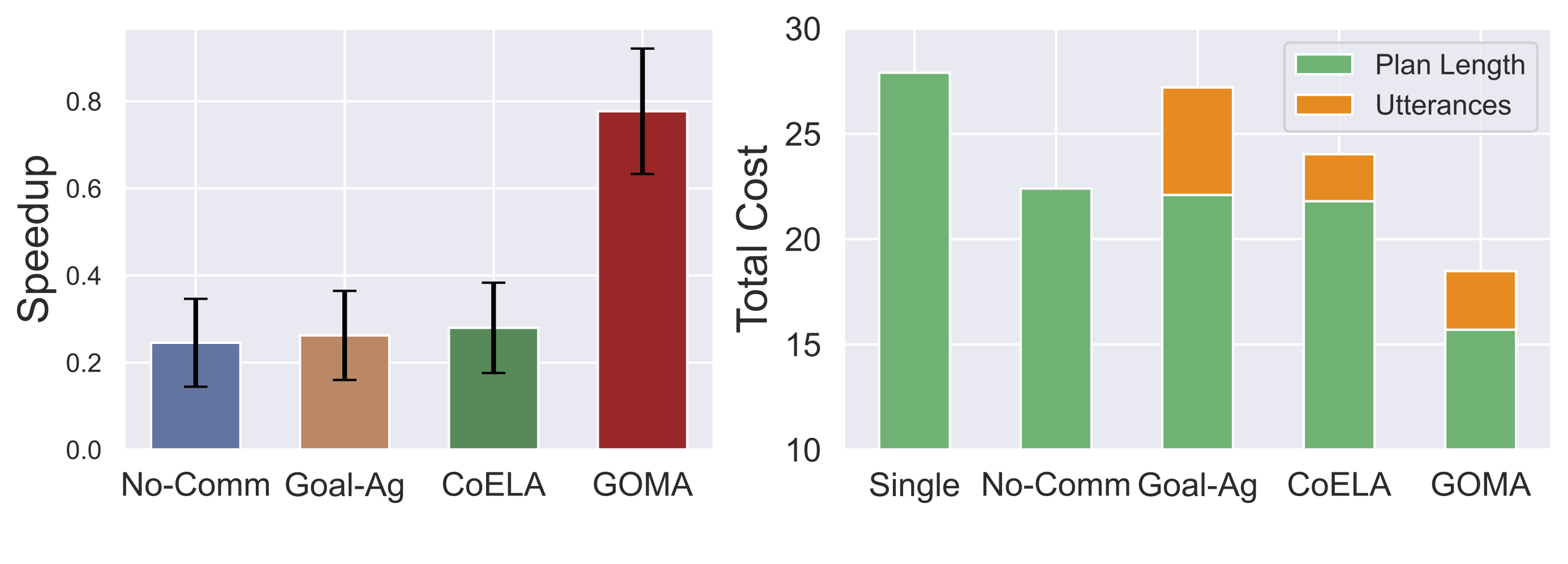}

    \end{subfigure}
    ~ 
    \begin{subfigure}[t]{0.48\textwidth}
    
        \centering
        \caption{Virtual-Home Human Rating}
        \includegraphics[width = \textwidth]{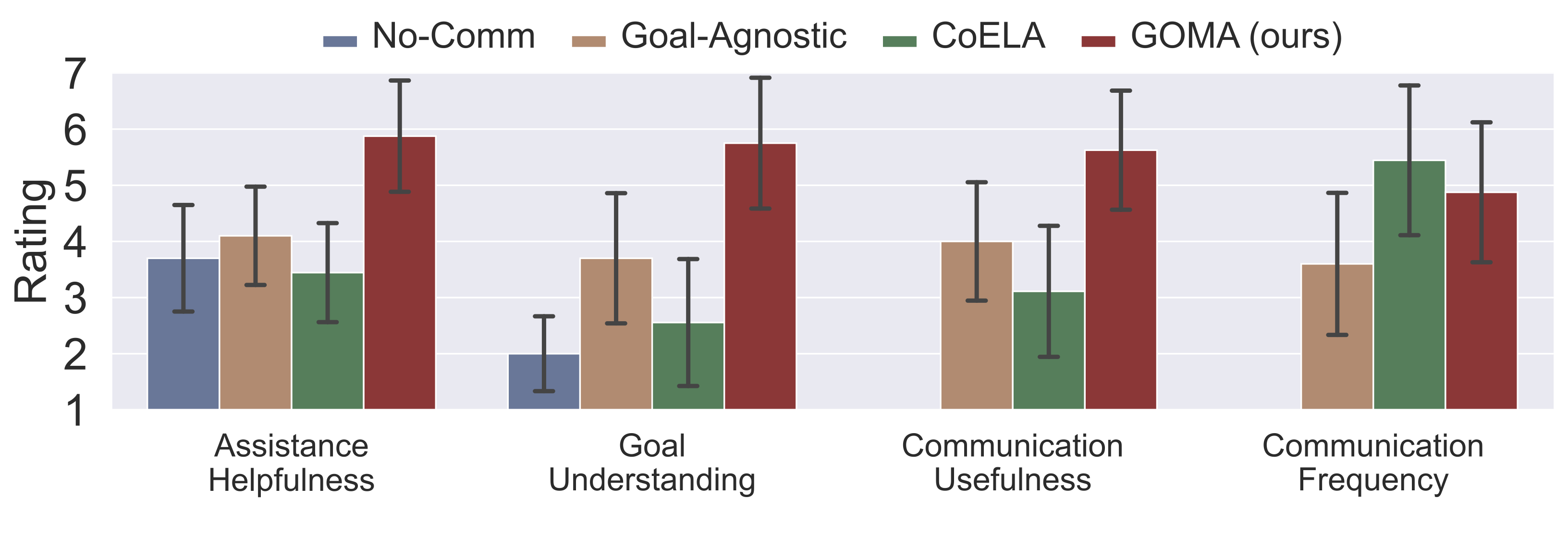}

    \end{subfigure}

    \vspace{-5pt}
    \caption{Experimental results in Overcooked and VirtualHome. The quantitative results from experiments (\textbf{a, b, c}) demonstrate that GOMA led to the greatest speedup (left) and least plan cost (right) compared to other baselines. In human subjective ratings (\textbf{d}), participants find GOMA to be more helpful and communicate more useful information than other models.}
    \label{fig:summary}
    \vspace{-10pt}
\end{figure*}

\begin{figure}[t!]
    \centering
    \includegraphics[width = 0.5\textwidth]{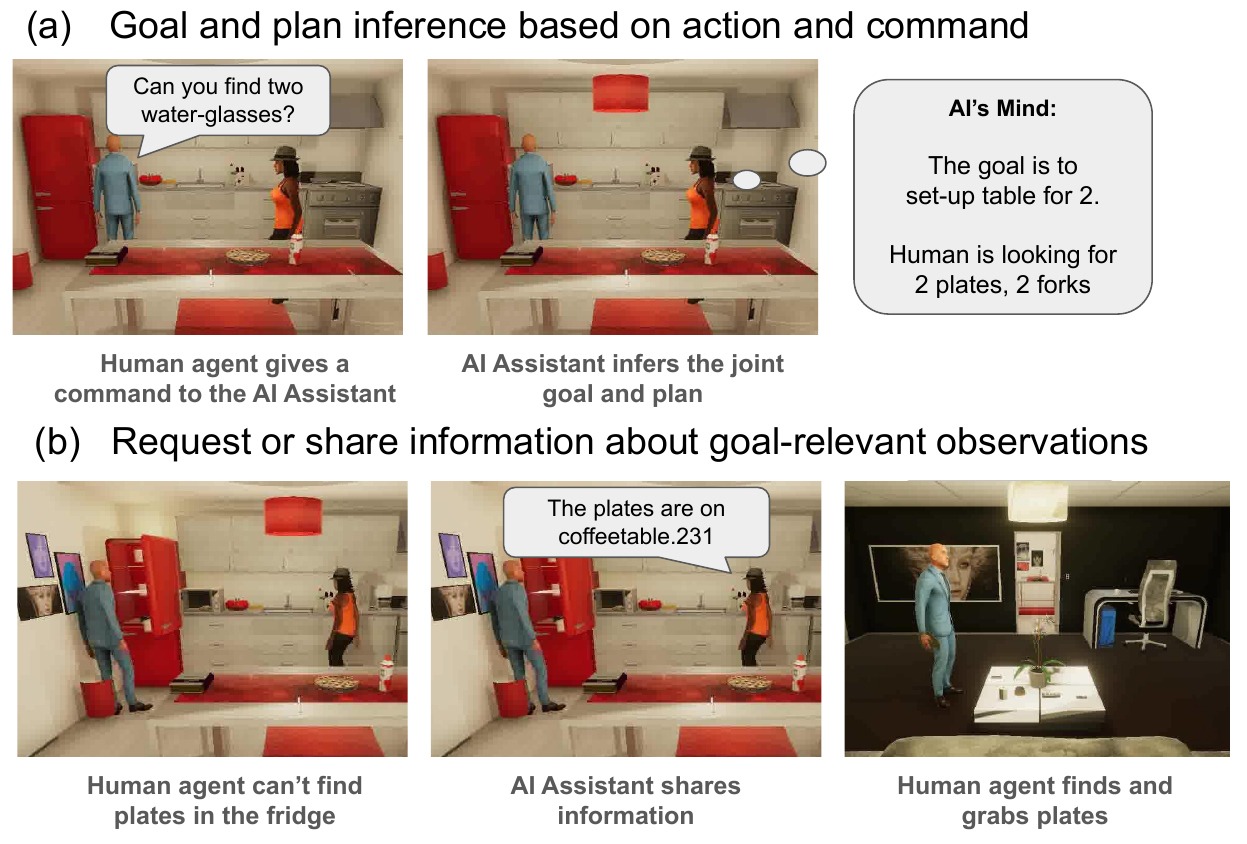}
    \caption{Example of typical communication enabled by GOMA in VirtualHome. (a) Once the human (in the blue shirt) gives a command to the AI Assistant (in the orange shirt), it infers the human goal and reasons that the human needs 2 plates and 2 forks. (b) As the AI watches the human agent opening the fridge, GOMA informs the human that the plates are on the coffee table. Consequently, the human goes to the coffee table to pick up a plate.}
    \label{fig:vh-storyboard}
    \vspace{-5pt}
\end{figure}

\begin{figure*}[t!]

    \centering
    \includegraphics[width = 0.9\textwidth]{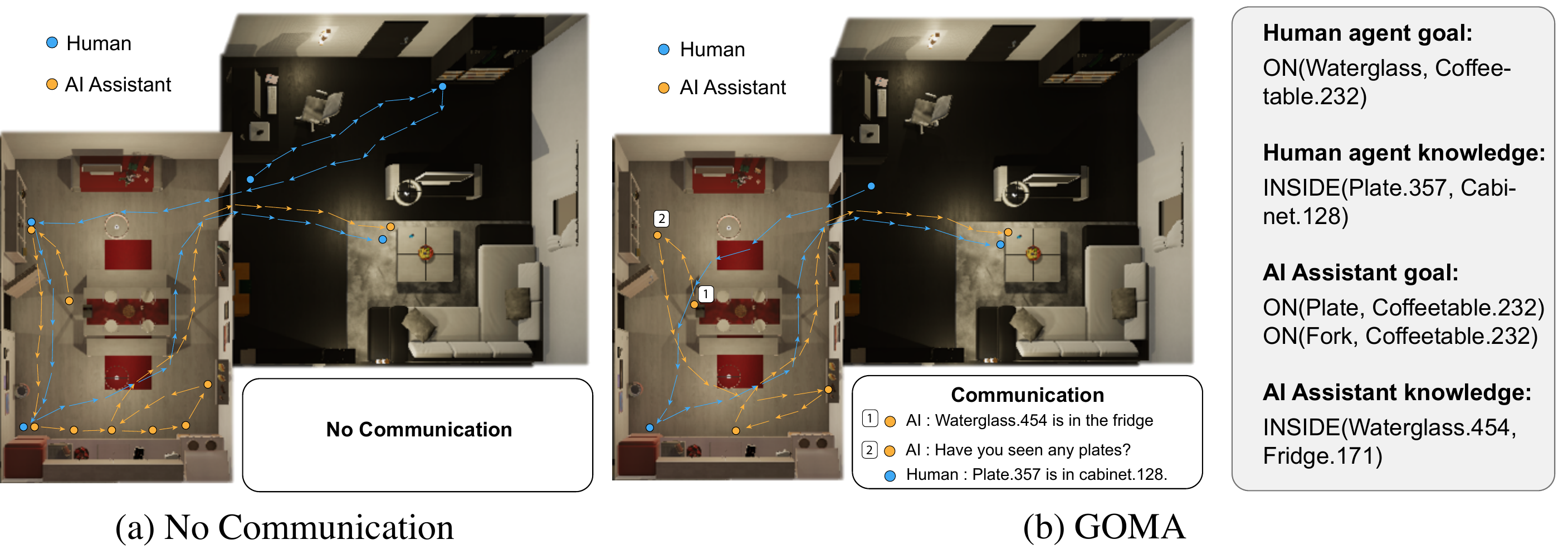}
    \caption{Agents' trajectories with No-Comm (left) and with GOMA (right) in a VirtualHome environment. In this example, the team goal is to set up a table for 1 person. The AI Assistant needs to find a plate and a fork while the human is looking for a water glass. Both agents have knowledge about the items that the other agent is looking for but not their own goal objects. In the No-Comm setting, the agents cannot share knowledge and have to open many containers to search for goal items. By inferring the other agent's goal and communicating goal-relevant knowledge, GOMA drastically reduces the total number of steps taken to complete the task.}
    \label{fig: vh-map}
    \vspace{-10pt}
\end{figure*}
\subsection{VirtualHome}
VirtualHome \cite{puig2020watch} is a multiagent household simulator. In VirtualHome, agents collaborate to complete daily household tasks. In our experiments, we included four common types of household tasks: Set Table, Load Dishwasher, Get Snacks, Stock Fridge. The goal for each task is defined as a set of goal predicates and their counts as defined in Table~\ref{tab:vh-goal}. The goal space was defined a priori and accessible to all models. In VirtulHome, each object is associated with a unique object ID, which we used in agents' communication to distinguish the referent from others (e.g. \textit{cabinet.145}).

\textbf{Simulation Experiment.}
We simulated 25 collaborative scenarios in VirtualHome across 4 goal types and 5 simulated apartments. Each episode was run 3 times and we reported the averaged results. We simulated the human agent using the MCTS planner from \cite{puig2020watch}. The simulated human agent requests help by sampling a subset of the goal predicates and replies to the AI assistant's questions. We compared our proposed method against four baselines: Single-agent, No-Communication (No-Comm), Goal-Agnostic (Goal-Ag), and LLM agent (CoELA). The first two were identical to the ones in Overcooked. The Goal-Ag baseline did not infer the joint goals and plan and instead randomly shared information about any objects that the human didn't know. We used CoELA \cite{zhang2023building} for the LLM agent, which had previously achieved great performance on human-AI collaborative tasks in VirtualHome. 

\textbf{Human Experiment.}
We developed an online human interface and conducted an experiment with 12 human participants recruited from 3 universities. We adopted a within-subject design where each participant completed 5 trials, each with a different AI assistant model (Single-Agent, No-Comm, Goal-Ag, CoELA, and GOMA) in a randomized order. In total the participants completed 60 trials over 20 unique task scenarios with each scenario receiving 3 datapoints. In all collaborative conditions, the interface included a chatbox that allowed the participant and the AI agent to send messages to each other. After completing a trial with an AI assistant, the participants were asked to rate the AI assistant based on four criteria: 1) the assistant is helpful; 2) the assistant understands your goal; 3) the assistant's communication is useful; and 4) the assistant communicates more than necessary. Each criterion is rated on a 7-point Likert scale (1 = Strongly Disagree, 7 = Strongly Agree).

\textbf{Metrics.}
In line with previous studies on VirtualHome \cite{puig2020watch}, we evaluated the models' performance by computing 1) speedups: counting the number of steps taken to complete the task, and 2) total costs: an overall cost metric that sums up the action and communication cost over the episode. 

\section{Results}

\subsection{Simulation Experiment}
The simulation results are shown in Fig.~\ref{fig:summary}\textbf{ab}. The advantage of collaboration is evident as the Single-agent baseline performed significantly worse than all other collaborative models. Overall, we find that across both Overcooked and Virtual-Home experiments, our model outperformed other baselines in all metrics. The differences between GOMA and other baselines are all statistically significant with $p< 0.01$ across two performance metrics.

In Overcooked, GOMA took on average 46.76 steps to complete the task, achieving a 44.61\% speedup. Our model completed the tasks with the lowest costs (M = 58.06) compared to the Heuristic-based model (M = 72.0) and No-Comm baseline (M = 65.05). Additionally, GOMA delivered the dishes in the best condition among all tested models, as signaled by the lowest coldness penalty (7.85).

In VirtualHome, GOMA took on average 20.08 steps to complete the task with a 55.8\% speedup. Despite having observed objects relevant to the human agent's goal, CoELA made few utterances (Mean = 3.03) and focused exclusively on communicating observations of its own goal. For example, when given a command \textit{"Please help me find a fork."}, CoELA would respond later \textit{"I found fork 323 in cabinet 132."} and did not share any knowledge that may be useful for the human's subgoal and plan. The Goal-Agnostic model makes frequent (Mean = 5.41) but mostly irrelevant utterances about possible goal objects. However, it did perform slightly better than the No-Comm baseline because, with enough utterances, it occasionally mentions useful information to the human agent.

Unlike baselines, GOMA can communicate and inquire about useful goal-relevant information with the human, leading to improved team performance. We include two qualitative examples of GOMA in VirtualHome simulations in Fig.~\ref{fig:vh-storyboard} and \ref{fig: vh-map}. In these examples, we show that due to partial observability, the AI Assistant and the human have exclusive knowledge about certain objects relevant to other agent's subgoals. GOMA allows the AI Assistant to inquire and inform another agent about this goal-relevant information. As a result, the agents can find the goal objects quickly without exhaustively opening and checking all containers.

\subsection{Human Experiment}
The human experiment results are shown in Fig.~\ref{fig:summary}\textbf{cd}. Similar to the simulation results, our proposed method had the greatest speedup over a single agent and outperformed all baselines in terms of plan costs. In contrast to the simulation study, the Goal-agnostic model here performed no better than No-Comm and CoELA as participants stopped paying attention to the assistant after it made too many statements irrelevant to the goal. This is shown in the participants' subjective ratings where participants reported that the Goal-Agnostic baseline communicated more than necessary.

The participants gave a higher subjective rating to our model than other baselines on all 4 items. Interestingly, even though CoELA and GOMA performed goal inference with the same method, the participants thought that only GOMA understood the human's goal. This is because by communicating goal-relevant information, GOMA implicitly expressed its understanding of the user's goal, whereas CoELA only communicated the progress of its own subgoal.

\section{Conclusion}

In this paper, we introduce GOMA, which enables an embodied AI assistant to efficiently and effectively communicate with a human user to achieve optimal cooperation. GOMA achieves this by reasoning about the other agent's mental state, assessing the misalignment between mental states, and then proactively initiating necessary communication to exchange goal-relevant information. Our experiments in Overcooked and VirtualHome demonstrate that embodied AI assistants built with GOMA can not only help achieve the human goal faster with lower total plan cost but also receive higher subjective ratings from human participants.

Our study is not without limitations. We have not evaluated GOMA on real-world robot assistants, which we intend to study in the future. We also plan to enhance the flexibility of the communication generation, so that it can communicate about \textit{any} information relevant to the task in an open-ended manner. Finally, we also aim to investigate more general belief representations that go beyond object states.

\section{Acknowledgement}

This project is funded in part by Lockheed Martin, DARPA Machine Common Science, and Schmidt AI 2050. We thank Hao Liu and Shitian Yang for their contribution.

\bibliographystyle{IEEEtran}

\bibliography{custom}

\end{document}